\documentstyle[11pt]{article}
\normalbaselines

 \newcommand{\cf}[1]{\mbox{\boldmath${#1}$}}

\begin{document}
\bibliographystyle{unsrt}
\title{Energy-sensitive and ``classical-like'' distances
between quantum states}
\vspace{1.0in}
\author{ V. V. Dodonov\thanks{e-mail: vdodonov@power.ufscar.br}\\
Departamento de F\'{\i}sica, Universidade Federal de
S\~ao Carlos,\\ Via Washington Luiz, km 235, 13565-905  S\~ao Carlos,
SP, Brasil\\[4mm]
O. V. Man'ko, V. I. Man'ko\thanks{e-mail: manko@na.infn.it}\\
P.N. Lebedev Physical Institute, Leninskii Prospekt 53, 117924 Moscow,
Russia\\[4mm]
A. W\"unsche\thanks{e-mail: awunsche@physik.hu-berlin.de}\\
Arbeitsgruppe ``Nichtklassische Strahlung,'' der Max-Planck-Gesellschaft,\\
 Humboldt University, Berlin, Germany}
 \maketitle

\begin{abstract}
We introduce the concept of the ``polarized'' distance, which
distinguishes the orthogonal states with different energies.
We also give new inequalities for the known Hilbert-Schmidt distance
between neighbouring states and express this distance in terms of the
quasiprobability distributions and the normally ordered moments.
Besides, we discuss the distance problem in the framework of the
recently proposed ``classical-like'' formulation of quantum mechanics, based
on the symplectic tomography scheme.
The examples of the Fock, coherent, ``Schr\"odinger cats,'' squeezed,
phase, and thermal states are considered.
\end{abstract}

\vspace{5mm}
{PACS Ref: 03.65.Bz, 42.50.Dv}

%

\newcommand{\be}{\begin{equation}}
\newcommand{\ee}{\end{equation}}
\newcommand{\br}{\begin{eqnarray}}
\newcommand{\er}{\end{eqnarray}}
\newcommand{\ds}{\displaystyle}
\newcommand{\bdm}{\begin{displaymath}}
\newcommand{\edm}{\end{displaymath}}
\def\im{\mbox{$\dot{\imath}$}}

\newpage

\section{Introduction}

Last years, an increasing interest to the problem
of distance between quantum states is observed.
Different motivations of this activity can be found in such fields as
quantum cryptography, quantum communications, or quantum computing.
Here we discuss the topic mainly from the point of view of quantum optics.
In view of recent impressive progress in creating and detecting various
types of nonclassical states of light or cooled particles in electromagnetic
traps, the problem of measures of distinguishability or closeness between
different quantum states becomes actual.
        For example, in quantum optics, the Glauber coherent states
\cite{glauber63}
\begin{equation}\label{def-coh}
| \alpha \rangle =\exp \left(-|\alpha |^2/2\right )
\sum _{n=0}^\infty \frac {\alpha ^n}{\sqrt{ n!}}| n\rangle
\end{equation}
are considered frequently as reference states ($|n\rangle$ means the Fock
state with the definite number of photons), so that the (pure) states
different
from (\ref{def-coh}) are called sometimes as {\it nonclassical states\/}.
But what is the quantitative measure of the ``nonclassicality?''
The simplest option is to use the so-called Mandel's parameter,
${\cal Q}=\overline{n^2}/\bar{n}\, -\bar{n} -1$,
which equals zero for all coherent states, since they have the Poissonian
photon statistics.
However, this parameter is adequate for a limited class of states.
Consider, for instance,
the {\it even and odd coherent states\/} introduced in \cite{DMM-74}
\begin{equation}
|\alpha;\pm\rangle=
\left(2\left[1\pm \exp(-2|\alpha|^2)\right]\right)^{-1/2}
\left(| \alpha\rangle \pm | -\alpha\rangle\right).
 \label{ev-od}
 \end{equation}
In this case, Mandel's parameter equals
${\cal Q}^{(\pm)}=\pm2|\alpha|^2/\sinh(2|\alpha|^2)$, and it
shows distinctly the qualitative difference between the states
$|\alpha\rangle$, $|\alpha;+\rangle$, and $|\alpha;-\rangle$,
but only for small values of $|\alpha|$. If $|\alpha|\gg 1$, then
${\cal Q}^{(\pm)}\approx 0$, although the states $|\alpha;\pm\rangle$
are still quite different from the coherent state. Moreover, for
 {\it generalized coherent states\/} \cite{titgla,zofia}
\begin{equation}\label{titul}
| \widetilde{\alpha }\rangle =\exp \left(-|\alpha |^2/2\right )
\sum _{n=0}^\infty \frac {\alpha ^n}{\sqrt{ n!}}\,\exp \left [i\varphi
\left(n\right )\right ]| n\rangle
\end{equation}
 we have identically ${\cal Q}\equiv 0$ for any function
$\varphi(n)$,
although the state $| \widetilde{\alpha }\rangle$ may be essentially
different from the Glauber state $| \alpha \rangle$.
For example, the choice
$\varphi(2k)=0\, (\mbox{mod}2\pi)$,
$\varphi(2k+1)=-\pi/2\, (\mbox{mod}2\pi)$
gives the so-called Yurke-Stoler state \cite{YuS}
\begin{equation}
|\widetilde{\alpha}\rangle_{YS}=
e^{-i\pi/4}\left(| \alpha\rangle +i | -\alpha\rangle\right)/\sqrt2
\label{Yur}
\end{equation}
which is considered, equally with the even and odd states, as
a representattive of a large family of ``Schr\"odinger cat states.''

The concept of distance gives a possibility to characterize more precisely
the neighbourhood or similarity between the quantum states.
However, the existing approaches
(see section~\ref{revdist}) seem
to suffer from certain drawbacks. Some of the available definitions of a
distance
are too complicated to perform concrete calculations. On the other hand,
some consequences of the traditional approaches, being correct
mathematically, contradict the physical intuition.
For example, the known definitions yield the same, at once, maximum 
possible value of the distance between any two orthogonal pure states, 
whereas from the physical point of view, the distance between the
first and the $100$th Fock states seems to be much greater
than that between, say, the $100$th and the $101$th states. The distance 
measures based on the density operators alone are
not sensible to the difference in energies.

In the present paper, we propose new measures which distinguish
different orthogonal states and which are simple enough to perform the
calculations, at least for the most important families of states used in
quantum optics. In our approach, the distance depends not only on the
density operators alone, but also on some extra fixed
positively definite operator.
Of course, following this way we meet the problem of the nonuniqueness
in the choice of this additional ``polarization'' operator.
Nonetheless, such a nonuniqueness seems not crucial
in many physical applications, where the special role of some operators
(like the Hamiltonian or the quantum number operator) is evident from the
beginning.
Another goal is to provide an analysis of the distance problem
in terms of the quasiprobability distributions and in the framework of
the ``classical-like'' formulation of quantum mechanics
proposed recently in \cite{mancini3-mancini4}.

The paper is organized as follows.
In section~\ref{revdist}, we give a review of the existing approaches
to the quantum distance problem.
In section~\ref{Hilb}, we concentrate on the properties of the
Hilbert-Schmidt distance (HSD)
and we express it in terms of the quasiprobability
functions and ordered moments.
In section~\ref{sec-ener-rho},
we propose several definitions of the energy-sensitive
distance in terms of the statistical operators (density matrices).
In section~\ref{sec-examples}, the
distinctions between different definitions are illustrated by examples of
the Fock, coherent, ``Schr\"odinger cat,'' squeezed, phase,
and thermal states.
The ``classical-like'' distances between quantum states are considered
in section~\ref{dist-clas}.
The last section contains brief conclusions.

\section{Previous approaches to the quantum distance problem}\label{revdist}
\setcounter{equation}{0}

The distance between two objects $a$ and $b$ is defined usually as a scalar
real function satisfying the following properties:
\begin{eqnarray}
\mbox{(I)}&&\; d(a,a)=0, \qquad d(a,b)>0, \quad \mbox{if } a\neq b,
\label{I}\\
\mbox{(II)}&&\; d(a,b)=d(b,a),
\label{II}\\
\mbox{(III)}&&\;  d(a,b)+d(b,c)\geq d(a,c)\,.
\label{treug}\end{eqnarray}
The property (III)  has a clear geometrical
meaning as the {\it triangle inequality\/}, and it implies
rather strong limitations on the possible choice of the function $d(a,b)$.
If the ``objects'' $a$ and $b$ are different pure quantum states, then the
distance must be some functional written
in terms of the Hilbert space vectors, $|a\rangle$ and $|b\rangle$,
representing the states. One should remember, however, that
the set of quantum states is in one-to-one correspondence not with the
whole Hilbert space of the wave functions, but with its projective factor
space, since the vectors $|\psi\rangle$ and
$e^{i\varphi}|\psi\rangle$ describe the same state.
All the requirements are satisfied, e.g., for the
{\it Fubiny-Study distance\/} \cite{Barg,Baltz,Anand91}
\begin{equation}
d^{(FS)}(\psi_1,\psi_2)=
\sqrt2\left(1-\left|\langle\psi_1|\psi_2\rangle\right|^2\right)^{1/2}
\label{FubStu}\end{equation}
(sometimes the factor $\sqrt2$ is replaced by
$1$ or $2$),
although a slightly different definition
\begin{equation}
d^{(\min)}(\psi_1,\psi_2)=
\inf_{\varphi}
\left\Vert|\psi_1\rangle - e^{i\varphi}|\psi_2\rangle\right\Vert=
\sqrt{2}\left(1-\left|\langle\psi_1|\psi_2\rangle\right|\right)^{1/2}
\label{primit}\end{equation}
is also possible \cite{Pati91}.
Taking a one-parameter
 family of states $\psi(t)$ generated by the time evolution operator,
 one obtains, both from (\ref{FubStu}) and (\ref{primit}), the infinitesimal
 distance along the evolution curve in the projective Hilbert space
\begin{equation}
 ds=\sqrt{2-2\left|\langle\psi(t)|\psi(t+dt)\rangle\right|^2}
\approx 2\sqrt{1-\left|\langle\psi(t)|\psi(t+dt)\rangle\right|}.
\label{infdist}
\end{equation}
The definition (\ref{infdist}) was used in studies devoted to
the geometrical
aspects of the quantum evolution and generalizations of the
time-energy uncertainty relations
\cite{Anand91,Pati91,AnAh90,Mont,Pati92,Grig,Hub93-1,HirHam,BraMil}.
For a family of states $\psi({\cf s})$ dependent on a
continuous vector parameter ${\cf s}=\left(s_1,s_2,\ldots,s_n\right)
\in {\bf R}^n$, one can introduce the Riemannian metrics according to
${\Vert\psi({\cf s}+d{\cf s}) -\psi({\cf s})\Vert}^2=
\gamma_{ij}ds_i ds_j$
and measure not the ``shortest'' distance (\ref{d-us}),
but the distance along a geodesics on a curved manifold, which can be
much greater than the ``shortest'' one. The concrete examples of the
geometries on the manifolds corresponding to the most known continuous
families of quantum states (namely, coherent, squeezed, and displaced
states) were studied in detail in
\cite{HirHam,Provost,Page,Anan90,Trif,Abe}.

Wootters \cite{Woot} proposed the distance between the pure states in the
form of the angle between the corresponding rays in the Hilbert space
$
d^{(W)}\left(|\psi_1\rangle,|\psi_2\rangle \right)=
\cos^{-1}\left|\langle\psi_2|\psi_1\rangle\right|
$.
For infinitesimaly close states,
the differential form of this distance coincides (up to a coefficient)
with (\ref{infdist}) \cite{Braun}. Recently, the Wootters and Fubini-Study
metrics were compared in \cite{Plast}.

Now let us turn to the {\it mixed\/} quantum
states, described by positively definite statistical operators
$\hat\rho$ with the unit trace: $\mbox{Tr}\hat\rho=1$.
The first definition of the distance between mixed states
in the physical literature, perhaps, was given in
 \cite{Jauch}
\begin{equation}
d^{(JMG)}\left(\hat{\rho}_1,\hat{\rho}_2 \right)
= \sup_{\Vert A\Vert=1}\left|\mbox{Tr}\left(\left[
\hat{\rho}_1-\hat{\rho}_2 \right]\hat A\right)\right|.
\label{d-jauch}
\end{equation}
Restricting the family of the bounded operators $\hat A$ in this definition
by the projection operators $\hat E=\hat E^2$, one obtains an equivalent
definition \cite{Dieks}
\begin{equation}
d^{(JMG)}\left(\hat{\rho}_1,\hat{\rho}_2 \right)
= \sup_{E}\left|\mbox{Tr}\left(\left[
\hat{\rho}_1-\hat{\rho}_2 \right]\hat E\right)\right|
= \frac12\left\Vert\hat{\rho}_1-\hat{\rho}_2\right\Vert_1,
\label{d-jauch-dieks}
\end{equation}
where $\left\Vert\hat{A}\right\Vert_1\equiv
\mbox{Tr}\sqrt{\hat{A}^{\dag}\hat{A}}\equiv
\sum|\lambda_n|$, the
summation being performed over all the eigenvalues $\lambda_n$ of the
operator $\hat{A}$.
Actually, the right-hand side of equation (\ref{d-jauch-dieks})
was used by Hillery \cite{Hil} as a starting point
in his definition of the distance between a state $\hat{\rho}$ and a given
family of ``classical'' states $\hat{\rho}_{cl}$ as
$\delta=\inf_{\rho_{cl}}\left\Vert\hat{\rho}
-\hat{\rho}_{cl}\right\Vert_1 $.
More sophisticated definitions of the distance were given, e.g., 
in~\cite{Ruch,Guz}. However, they are so complicated from the point of view
of calculations, that no explicit examples were considered.

One of the most frequently cited in the physical literature definitions is
the so-called {\it Bures-Uhlmann distance\/} (BU-distance)
\cite{Bures,Uhlm}. It has the
form (see also \cite{Dieks,Gud,Hub,Joz})
\begin{equation}
d^{(BU)}\left(\hat{\rho}_1,\hat{\rho}_2 \right)
=\left( 2-2\mbox{Tr}\sqrt{
\hat{\rho}_1^{1/2}\hat{\rho}_2 \hat{\rho}_1^{1/2}}\right)^{1/2},
\label{d-buruhl}
\end{equation}
where the operator $\hat{\rho}^{1/2}$ is defined as the
{\it positively semidefinite Hermitian operator\/}
satisfying the relation $\left(\hat\rho^{1/2}\right)^2=\hat\rho$.
This operator is unique.
Although the right-hand side of (\ref{d-buruhl}) seems asymmetrical with
respect to $\hat{\rho}_1$ and $\hat{\rho}_2$, actually
$d^{(BU)}\left(\hat{\rho}_1,\hat{\rho}_2 \right)=
d^{(BU)}\left(\hat{\rho}_2,\hat{\rho}_1 \right)$ \cite{Joz}.
For {\it pure\/} quantum states $\hat\rho_{\psi}=|\psi\rangle\langle\psi|$,
the BU-distance coincides with
 the ``minimal'' distance (\ref{primit}) due to the relations
$\hat\rho_{\psi}^{1/2}=\hat\rho_{\psi}^{2}=\hat\rho_{\psi}$.
If one of the states is pure, then
\begin{equation}
d^{(BU)}\left(|\psi\rangle\langle\psi|,\hat{\rho} \right)=
\sqrt2\left(1-\sqrt{\langle\psi|\hat{\rho}|\psi\rangle}\right)^{1/2}.
\label{BUpsirho}
\end{equation}
However, the calculations are much more involved
in the generic case of nondiagonal statistical operators, so that
the explicit forms of the Bures-Uhlmann distance were
found only for finite-dimensional $N$$\times $$N$ density matrices
(especially, for $N=2$ and $N=3$) \cite{Hub,Hub93-2,Slat1} and recently for
squeezed thermal states \cite{Twam,Slat2} and displaced thermal states
\cite{scut}.

\section{Distances based on the Hilbert-Schmidt norm}\label{Hilb}
\setcounter{equation}{0}

A simple expression for the distance between quantum states,
enabling to perform calculations for the most important
classes of states (at least in the problems of quantum optics),
is based on the Hilbert-Schmidt norm
$||\hat{A}||_2 \equiv \sqrt{{\rm Tr}(\hat{A}^\dagger \hat{A})}$.
The Hilbert-Schmidt distance (HSD) of two statistical
operators $\hat{\rho}_1$ and $\hat{\rho}_2$ is defined as
\cite{Baltz,Anand91,HirHam,Dieks,Sam,Wun,Orl}
\begin{eqnarray}
d^{(HS)}(\hat{\rho}_1,\hat{\rho}_2)&=&
\left\Vert\hat{\rho}_1-\hat{\rho}_2\right\Vert_2
=\left\{{\rm Tr}\left[(\hat{\rho}_1-\hat{\rho}_2)^2\right]\right\}^{1/2}
\nonumber\\
&=&\left[{\rm Tr}\left(\hat{\rho}_1^2\right)+
{\rm Tr}\left(\hat{\rho}_2^2\right) -2\,{\rm Tr}
\left(\hat{\rho}_1\hat{\rho}_2\right)\right]^{1/2}.
\label{d-us}
\end{eqnarray}
In particular
(we write simply $d$ instead of
$d^{(HS)}$ in all cases when it does not lead to a confusion),
\begin{equation}
d(|\psi\rangle \langle \psi|,\hat\rho) 
= \left[1+{\rm Tr}\left(\hat{\rho^2}\right)-
2\langle \psi|\hat{\rho}|\psi\rangle \right]^{1/2}
\le \sqrt2\left[1-\langle \psi|\hat{\rho}|\psi\rangle \right]^{1/2},
\label{puremix}
\end{equation}
so the HSD (\ref{d-us}) goes to the Fubini-Study distance (\ref{FubStu})
in the special case of two pure states.
The possible values of the Hilbert-Schmidt distance are restricted by
$0 \le d(\hat{\rho}_1,\hat{\rho}_2) \le \sqrt{2}$,
the maximum $\sqrt{2}$ being reached for any pair of orthogonal (pure)
states.

In many cases,
it is convenient to describe the quantum states with the aid
of quasiprobability distributions, which
can be written as special cases of the
general Cahill-Glauber $s$-distribution \cite{cahgla}
\begin{equation}
W(\alpha,s)=\mbox{Tr}\left[\hat\rho\hat{T}(\alpha,s)\right],
\label{defcahgla}
\end{equation}
where
\[
\hat{T}(\alpha,s)=\int \frac{d^2\zeta}{\pi}
\exp\left[\zeta\left(\hat{a}^{\dagger} -\alpha^*\right)
-\zeta^*\left(\hat{a} -\alpha\right) +\frac{s}{2}|\zeta|^2\right],
\]
 $\alpha,\zeta$ are complex numbers and $\hat{a},\hat{a}^{\dagger}$
are the boson annihilation and creation operators (in one dimension, for
simplicity). The choice $s=0$ (with $\alpha=(q+ip)/\sqrt2$) yields
the Wigner function \cite{wigner32}
$
W(q,p)\equiv\int du \exp\left(ipu\right) \langle q-u/2|
\hat{\rho}| q+ u/2\rangle
$.
For $s=-1$, we have the so-called Husimi-Kano or $Q$-function
\cite{husimi-kano}
$
W(\alpha,-1)\equiv Q(\alpha)=\langle\alpha|\hat\rho|\alpha\rangle
$,
whereas in the case $s=+1$ we arrive at the Glauber-Sudarshan function
$P(\alpha)\equiv W(\alpha,+1)$ which yields the ``diagonal'' representation
of the statistical operator \cite{glauber-sudarshan}
$
\hat\rho=\int P(\alpha)|\alpha\rangle\langle\alpha|d^2\alpha/\pi
$.
Using (\ref{defcahgla}) one can write the
Hilbert-Schmidt distance in terms of integrals over the phase space:
\begin{eqnarray}
d^2(\hat{\rho}_1,\hat{\rho}_2)&=&
\int \frac{dq\,dp}{2\pi}\left[W_1(q,p)-W_2(q,p)\right]^2
\label{dis-WW}\\
&=& \int \frac{d^2\alpha}{\pi}
\left[Q_1(\alpha)-Q_2(\alpha)\right]
\left[P_1(\alpha)-P_2(\alpha)\right]
\label{dis-QP}\\
&=&\int \frac{d^2\alpha}{\pi}\frac{ d^2\beta}{\pi}
e^{-|\alpha-\beta|^2}
\left[P_1(\alpha)-P_2(\alpha)\right]
\left[P_1(\beta)-P_2(\beta)\right].
\label{dis-PP}
\end{eqnarray}

If one knows (e.g., from experimental data) all
normally ordered moments
$
M^{(k,l)}={\rm Tr}\left(\hat{a}^{\dagger\,k}\hat{a}^l\hat{\rho}\right)
$,
then the statistical operator $\hat{\rho}$ can be reconstructed
as follows \cite{wu,wubu,wuensche}:
\begin{equation}
\hat{\rho}=\sum_{k=0}^\infty \sum_{l=0}^\infty M^{(k,l)} \hat{a}_{k,l}\;,
\qquad
\hat{a}_{k,l} \equiv \sum_{j=0}^{\min\{k,l\}}\frac{(-1)^j|l-j\rangle
\langle k-j|}{j!\sqrt{(k-j)!(l-j)!}}\,.
\label{akl}
\end{equation}
Using this formula
one can write the Hilbert-Schmidt distance in the form of a series
\begin{equation}
d^2(\hat{\rho}_1,\hat{\rho}_2)
 =\sum_{s=0}^{\infty}
\sum_{k=0}^s \sum_{l=0}^s \frac{(-1)^{s+k+l} s!}{k!(s-k)!l!(s-l)!}
\Delta M^{(k,l)}\Delta M^{(s-k,s-l)}\,,
\label{dis-mom}
\end{equation}
where
$\Delta M^{(k,l)}\equiv M^{(k,l)}_1 -M^{(k,l)}_2$.
For example, in the case of the coherent state $|\alpha\rangle$
one has $M^{(k,l)}=\alpha^{*\,k}\alpha^l$ and
(\ref{dis-mom}) converges to the closed expression (\ref{coh-coh}).

An advantage of the Hilbert-Schmidt distance is that it permits to obtain
simple inequalities for the distances between neighbouring states.
Consider, for example, the distance between an arbitrary state
$\hat{\rho}$ and the vacuum state $|0\rangle \langle 0|$.
Using formula (\ref{puremix}) and the identities
$
\sum \langle n|\hat{\rho}|n\rangle\equiv 1$,
$\sum n \langle n|\hat{\rho}|n\rangle\equiv\overline{n}$,
one can write the following chain of relations:
\begin{eqnarray}
d(\hat{\rho},|0\rangle \langle 0|)
&\le& \left[2(1-\langle 0|\hat{\rho}|0\rangle)\right]^{1/2}
= \left[2\sum_{n=1}^\infty \langle n|\hat{\rho}
|n\rangle\right]^{1/2} \nonumber\\
&\le&  \left[2\sum_{n=1}^\infty n\langle n|\hat{\rho}|n\rangle\right]^{1/2}
= \sqrt{2\overline{n}}.
\label{bound-0}
\end{eqnarray}
This inequality is useful if $\overline{n}\ll 1$.
For an arbitrary reference Fock state $|n\rangle \langle n|$,
one can prove in a similar way the inequalities
\begin{equation}
d\left(\hat{\rho},|n\rangle\langle n| \right)
\le\sqrt2\left[\langle 0|\bar{\rho}|0\rangle +
\overline{n}- n\langle n|\hat{\rho}|n\rangle\right]^{1/2},
\label{bound-n}
\end{equation}
\begin{equation}
d(\hat{\rho},|n\rangle \langle n|) \le \sqrt2\left[
\sigma_n + \left(n-\overline{n}\right)^2\right]^{1/2},
\label{vardif}
\end{equation}
where $\sigma_n\equiv \overline{n^2} -\left(\overline{n}\right)^2$ is the
variance of the number operator in the state $\hat{\rho}$.

In general, one can identify the quantum state not necessarily with the
statistical operator $\hat\rho$, but with any function of this operator
$f(\hat\rho)$. As a consequence, a whole family of the modified
Hilbert-Schmidt distances can be introduced according to the definition
\begin{eqnarray}
\Delta_f(\hat{\rho}_1,\hat{\rho}_2)&=&
\left\Vert f(\hat{\rho}_1)- f(\hat{\rho}_2)\right\Vert_2
=\left({\rm Tr}\left\{\left[f(\hat{\rho}_1)-f(\hat{\rho}_2)\right]^2\right\}
\right)^{1/2}
\nonumber\\
&=&\left({\rm Tr}\left[f^2(\hat{\rho}_1)\right]+
{\rm Tr}\left[f^2(\hat{\rho}_2)\right]
 -2\,{\rm Tr}\left[f(\hat{\rho}_1)f(\hat{\rho}_2)\right] \right)^{1/2}.
\label{d-f}
\end{eqnarray}
For pure states, $\Delta_f\,$-distances coincide with the Fubini-Study
distance
(\ref{FubStu}) for any reasonable function $f(\hat{\rho})$.
However, for mixed states the new distances are essentially different.
For example,
choosing $f(\hat\rho)=\hat\rho^{1/2}$ we obtain the distance
\begin{equation}
\tilde d\left(\hat{\rho}_1,\hat{\rho}_2 \right)=
\left[2-2\mbox{Tr}\left(
\hat{\rho}_1^{1/2}\hat{\rho}_2^{1/2}
\right)\right]^{1/2},
\label{d-root1}
\end{equation}
which coincides with
the Bures-Uhlmann distance (\ref{d-buruhl}) for any {\it commuting\/}
operators $\hat\rho_1$ and $\hat\rho_2$ (remember that the pure state
projection operators $|\psi\rangle\langle\psi|$ and
$|\varphi\rangle\langle\varphi|$ do not commute if
$|\psi\rangle\neq |\varphi\rangle$). If one of the states is pure, then
\begin{equation}
\tilde d(|\psi\rangle \langle \psi|,\hat\rho) =
 \sqrt2\left[1-\langle \psi|\hat{\rho}|\psi\rangle \right]^{1/2}
\label{puremix-1/2},
\end{equation}
so the inequalities (\ref{bound-0})-(\ref{vardif}) hold for the
$\tilde d$-distance, as well.

\section{Energy-sensitive distance between quantum states}
\label{sec-ener-rho}
\setcounter{equation}{0}

The Hilbert-Schmidt distance between any states cannot exceed the limit
value $\sqrt2$. In principle, one could ``stretch'' the distance between
remote states, introducing some monotonous function $F(d)$ with the
property $F(\sqrt2)=\infty$. But such a simple modification
yields the same (although infinite)
distance for any pair of orthogonal states.

To distinguish orthogonal states with different sets
of quantum numbers, we have to break the symmetry of the Hilbert space
with respect to ``rotations'' of the basis, i.e., to fix some ``direction''
given by a {\it positively definite Hermitian\/} ``reference'' operator
$\hat Z$.
However, we still want to use the advantage of the Hilbert-Schmidt norm.
So, we define the ``Z-polarized'' distance as
\begin{eqnarray}
d_Z\left(\hat{\rho}_1,\hat{\rho}_2 \right)&=&
\left\Vert\hat{Z}^{1/2}\left(\hat{\rho}_1 -\hat{\rho}_2
\right)\right\Vert_2=
\left[\mbox{Tr}\left(\hat{Z}
\left[\hat{\rho}_1 -\hat{\rho}_2\right]^2
\right)\right]^{1/2}
\nonumber\\&=&
\left[\mbox{Tr}\left(\hat{Z}
\left[\hat{\rho}_1^2 +\hat{\rho}_2^2 -
\hat{\rho}_1\hat{\rho}_2 -\hat{\rho}_2\hat{\rho}_1
\right]\right)\right]^{1/2} .
\label{d-Zlob}
\end{eqnarray}
Another possible definition is
\begin{eqnarray}
\tilde{d}_Z\left(\hat{\rho}_1,\hat{\rho}_2 \right)&=&
\left\Vert\hat{Z}^{1/2}\left(\hat{\rho}_1^{1/2}-\hat{\rho}_2^{1/2}
\right)\right\Vert_2=
\left[\mbox{Tr}\left(\hat{Z}
\left[\hat{\rho}_1^{1/2}-\hat{\rho}_2^{1/2}\right]^2
\right)\right]^{1/2}
\nonumber\\&=&
\left[\mbox{Tr}\left(\hat{Z}
\left[\hat{\rho}_1+\hat{\rho}_2 -
\hat{\rho}_1^{1/2}\hat{\rho}_2^{1/2}-\hat{\rho}_2^{1/2}\hat{\rho}_1^{1/2}
\right]\right)\right]^{1/2} .
\label{tild-Zlob}
\end{eqnarray}
Evidently, both the definitions satisfy all the axioms
due to the properties of the Hilbert-Schmidt norm (since we simply apply this
norm to the ``scaled'' operators $\hat{Z}^{1/2}\hat{\rho}$ or
$\hat{Z}^{1/2}\hat{\rho}^{1/2}$).
In the special case of pure quantum states
$\hat{\rho_i}=|\psi_i\rangle\langle\psi_i|$, we have
\begin{eqnarray}
&&d_Z^2\left(|\psi_1\rangle,|\psi_2\rangle \right)
=\tilde{d}_Z^2\left(|\psi_1\rangle,|\psi_2\rangle \right)
=\langle\psi_1|\hat{Z} |\psi_1\rangle +
\langle\psi_2|\hat{Z} |\psi_2\rangle \nonumber\\ &&-
\langle\psi_1|\hat{Z} |\psi_2\rangle \langle\psi_2|\psi_1\rangle -
\langle\psi_2|\hat{Z} |\psi_1\rangle \langle\psi_1|\psi_2\rangle\,.
\label{d-Zpsi}
\end{eqnarray}
If $\hat{Z}$ coincides with the unity operator,  (\ref{d-Zpsi}) goes to
the Fubini-Study distance (\ref{FubStu}).

A possibility of using some extra operators to define
the distance was mentioned in study~\cite{Jauch}
whose authors considered the construction
$\mbox{Tr}\left(\hat{A}\left[\hat{\rho}_1-\hat{\rho}_2\right]\right)$.
However, it was rejected on the grounds of the unboundness,
if {\it all\/} observables
$A$ are admitted (the authors of \cite{Jauch} started from the rough
definition: ``Two states are
close to each other if all the expectation values of observables are close
to each other''). Here we {\it fix\/} the operator $\hat{Z}$, depending
on the concrete physical problem.

 In the case of quantum optics, a natural choice of $\hat{Z}$ is
the quantum number operator
\begin{equation}
\hat N = \hat a^{\dag} \hat a.
\label{defN}
\end{equation}
Then the $N$-distance between the Fock states $|n\rangle$ and $|m\rangle$
reads
\begin{equation}
d_N(|m\rangle,|n\rangle)=\left(1-\delta_{mn}\right)\sqrt{m +n}\,.
\label{dist-mn-N}
\end{equation}
We see that $d_N(|m\rangle,|0\rangle) >d_N(|n\rangle,|0\rangle)$ if $m>n$,
 i.e., higher the energy, more is the distance
from the ground state.
Nonetheless, the $N$-distance also does not seem to be ideal. Consider,
for instance, two Fock states with $m,n\gg 1$. Then
$d_N(|m\rangle,|n\rangle)\gg 1$,
even if $|m-n|\sim 1$. Such a property of the distance (\ref{d-Zlob})
does not agree completely with our intuition.
This drawback can be removed, if we assume the following definition:
\begin{equation}
D_Z^2\left(\hat{\rho}_1,\hat{\rho}_2 \right)
=\mbox{Tr}
\left(\Delta\hat{\rho}\hat Z
\Delta\hat{\rho} \right)
-\frac{\left[\mbox{Tr}
\left(\Delta\hat{\rho}\hat Z^{1/2} \Delta\hat{\rho} \right)\right]^2}
{\mbox{Tr}\left(\Delta\hat{\rho} \right)^2},
\label{d-new}
\end{equation}
where
$\Delta\hat{\rho}\equiv \hat{\rho}_1-\hat{\rho}_2$.
The right-hand side of Eq.~(\ref{d-new}) is nonnegative, since it can
be written as
\begin{equation}
D_Z^2=\mbox{Tr}\left(\Delta\hat{\rho}\right)^2
\left\langle \left( Z^{1/2} -
\left\langle Z^{1/2}\right\rangle \right)^2\right\rangle,
\label{d-new1}
\end{equation}
where the average value is defined as
$
\langle{ Z}\rangle \equiv
\mbox{Tr}\left(\Delta\hat{\rho}\hat Z \Delta\hat{\rho} \right)
/\mbox{Tr}\left(\Delta\hat{\rho} \right)^2
$.
We shall cautiously name $D_Z$ as a {\it quasidistance\/}, since
we have no proof of the triangle inequality for {\it any\/} states.
Applying (\ref{d-new}) with $\hat{Z}=\hat{N}$ to the Fock states,
we obtain
\begin{equation}
D_{ N}\left(|n\rangle,|m\rangle \right) =
\left|\sqrt{n}-\sqrt{m}\right|/\sqrt2 .
\label{dist-n*}
\end{equation}
This expression obviously satisfies the triangle inequality. Moreover,
it is in agreement with the representation of the Fock states in
the phase space as circles whose radii are proportional to the square
root of the energy \cite{QE,Sch}. In such a case, the distance
between the $100$th and $101$th states is less than that between
the ground and the first excited states.

A disadvantage of the definition (\ref{d-new}) is that it
complicates significantly calculations for non-Fock states.
In the case of coherent states, the calculations are simplified if one
slightly modifies the definition of the quasidistance in the following way:
\begin{equation}
\widetilde{D}_a^2\left(\hat{\rho}_1,\hat{\rho}_2 \right)
=
\mbox{Tr}
\left(\Delta\hat{\rho}\hat a^{\dagger}\hat a
\Delta\hat{\rho} \right)
-\frac{\left|\mbox{Tr}
\left(\Delta\hat{\rho}\hat a \Delta\hat{\rho} \right)\right|^2}
{\mbox{Tr}\left(\Delta\hat{\rho} \right)^2}.
\label{D-newtil}
\end{equation}
Then
\begin{equation}
\widetilde{D}_a\left(|\alpha\rangle,|\beta\rangle \right) = \frac1{\sqrt2}
|\alpha-\beta|\sqrt{1+\exp\left(-|\alpha-\beta|^2\right)}\,.
\label{dist-n-cohtil}
\end{equation}
The right-hand side of Eq.~(\ref{dist-n-cohtil}) is a monotonous function
of $|\alpha-\beta|$, increasing from $|\alpha-\beta|$ at
$|\alpha-\beta|\ll 1$ to $|\alpha-\beta|/\sqrt2$ at $|\alpha-\beta|\gg 1$.
Although we have no proof that the quasidistance $\widetilde{D}_a$
satisfies the triangle inequality (\ref{treug})  for {\it all\/} states,
we can prove that the function
(\ref{dist-n-cohtil}) satisfies this inequality
for all values of $\alpha$ and $\beta$.

\section{Examples}\label{sec-examples}
\setcounter{equation}{0}

\subsection{Coherent and Fock states}

For two {\it coherent states\/}
$|\alpha\rangle$ and $|\beta\rangle$, one finds
\begin{equation}
d(|\alpha\rangle , |\beta\rangle )=
\sqrt{2}\left[1-\exp\left(-|\alpha-\beta|^2\right)\right]^{1/2}.
\label{coh-coh}
\end{equation}
If $|\alpha-\beta| \ll 1 $, then
$d(|\alpha\rangle , |\beta\rangle )\approx \sqrt{2}\,|\alpha-\beta|$
is proportional to the geometric distance
of the displacement parameters $\alpha$ and $\beta$ in the complex plane,
but it goes to $\sqrt{2}$ when $|\alpha-\beta| \gg 1$.
The $N$-distance (\ref{d-Zpsi}) between
the coherent states is given by
\begin{equation}
d_N\left(|\alpha\rangle,|\beta\rangle \right) =
\left[|\alpha|^2+|\beta|^2 -2\,\mbox{Re}\left(\beta^*\alpha\right)
\exp\left(-|\alpha-\beta|^2\right)\right]^{1/2},
\label{dist-n-coh}
\end{equation}
so $d_N(|\alpha\rangle,|0\rangle) >d_N(|\beta\rangle,|0\rangle)$ if
$|\alpha|>|\beta|$.
The $N$-distance is equal to the geometrical distance
$|\alpha-\beta|$ in the complex
plane of parameters, if $\mbox{Re}\left(\alpha\beta^*\right)=0$
(i.e., for orthogonal directions in the complex plane).
In Fig.~1, we plot the HS- and $N$-distances
between the Fock state
$|m\rangle$ and the coherent state $|\alpha\rangle$
\begin{equation}
d^{(HS)}(|\alpha\rangle,|m\rangle)= \sqrt2\left(1-
\frac{|\alpha|^{2m}}{m!}e^{-|\alpha|^2}\right)^{1/2},
\label{dist-coh-fock}
\end{equation}
\begin{equation}
d_N(|\alpha\rangle,|m\rangle)= \left( m+ |\alpha|^2 -
\frac{2|\alpha|^{2m}}{(m-1)!}e^{-|\alpha|^2}\right)^{1/2},
\label{dist-n-coh-fock}
\end{equation}
as functions of the mean photon number $|\alpha|^2$ for fixed values
of $m=1,2,3$. The HS-distance has a minimum at $|\alpha|^2=m$.
For small values of $|\alpha|^2$, we have
$d^{(HS)}(|\alpha\rangle,|m\rangle) > d^{(HS)}(|\alpha\rangle,|n\rangle)$
if $m > n$, but this inequality changes its sign if $|\alpha|^2$ is
sufficiently large. The N-distance has a minimum only for $m=1$, and the
$m$-dependence is monotonous for all values of $|\alpha|^2$.

\subsection{Squeezed vacuum states}

The {\it squeezed vacuum state\/} \cite{squeez} depends on the
complex parameter $\zeta$ with $|\zeta|<1$
\begin{equation}
|\zeta\rangle
= \left(1-|\zeta|^2\right)^{1/4}\sum_{n=0}^{\infty}
\frac{\sqrt{(2n)!}}{2^n n!} \zeta^n
|2n\rangle.
\label{defsqueez}
\end{equation}
The HS-distance between the states $|\zeta_1\rangle$ and $|\zeta_2\rangle$
reads (see also \cite{wuensche,ManW})
\begin{equation}
d(|\zeta_1\rangle , |\zeta_2\rangle )
=\frac{\sqrt{2}\,|\zeta_1-\zeta_2|}{\left(|1-\zeta_1\zeta_2^*|
\left[|1-\zeta_1\zeta_2^*| +\sqrt{\left(1-|\zeta_1|^2\right)
\left(1-|\zeta_2|^2
\right)}\right]\right)^{1/2}}.
\label{sq-sq}
\end{equation}
For $|\zeta_1|\ll1$ and $|\zeta_2|\ll 1$, this is the geometric
distance of the complex squeezing parameters.
Using the parametrisation $\zeta=\tanh\tau\, e^{i\phi}$, $\tau\ge 0$,
we have a simplified formula in the case of $\phi_1=\phi_2$:
\begin{equation}
d\left(|\zeta_1\rangle,|\zeta_2\rangle \right) =
\frac{2\sinh\left[\frac12\left(\tau_1-\tau_2\right)\right]}
{\sqrt{\cosh\left(\tau_1-\tau_2\right)}}.
\label{squeez-d-par}
\end{equation}
For $\tau_2=0$, (\ref{squeez-d-par}) gives the distance between the vacuum
state and the squeezed state $|\zeta_1\rangle$.

 The $N$-distance can be expressed as
\begin{eqnarray}
d_{ N}^2\left(|\zeta_1\rangle,|\zeta_2\rangle \right) &=&
\frac{|\zeta_1|^2}{1-|\zeta_1|^2} + \frac{|\zeta_2|^2}{1-|\zeta_2|^2}
\nonumber\\
&+&2\frac{|\zeta_1\zeta_2|^2-\mbox{Re}\left(\zeta_1\zeta_2^*\right)}
{|1-\zeta_1\zeta_2^*|^3}
\sqrt{\left(1-|\zeta_1|^2\right)\left(1-|\zeta_2|^2\right)}\,.
\label{dist-squeez}
\end{eqnarray}
If $|\zeta_{1,2}|\ll 1$, then (\ref{dist-squeez})
has the same limit as the ``unpolarized'' Hilbert-Schmidt distance
(\ref{sq-sq}):
$d_N\approx d\approx \left|\zeta_1-\zeta_2\right|$.
However, for large values of the squeezing
parameter these two distances become completely different. For example,
in the special case $\Delta\phi\equiv \arg\zeta_1 -\arg\zeta_2=0$ we have
instead of (\ref{squeez-d-par}) the expression
($\tau_j\equiv|\zeta_j|$)
\begin{equation}
d_{ N}^2\left(|\zeta_1\rangle,|\zeta_2\rangle \right) =
\sinh^2\tau_1 +\sinh^2\tau_2 -
\frac{2\sinh\tau_1\sinh\tau_2}
{\cosh^2\left(\tau_1-\tau_2\right)}
\label{squeez-D-par}
\end{equation}
and $d_{ N}\left(|\zeta\rangle,|0\rangle \right) = \sinh\,\tau$.

\subsection{``Schr\"odinger cat'' states}

Now let us consider the family of the ``Schr\"odinger cat'' states
\begin{equation}
|\alpha;\varphi\rangle=
\left(2\left[1+\cos\varphi \exp(-2|\alpha|^2)\right]\right)^{-1/2}
\left(| \alpha\rangle +e^{i\varphi} | -\alpha\rangle\right).
 \label{cat}
 \end{equation}
The special cases of this family are even states ($\varphi=0$),
odd states ($\varphi=\pi$), and the Yurke-Stoler states ($\varphi=\pi/2$).
A more general set of states
$|\alpha;\tau,\varphi\rangle \sim
| \alpha\rangle +\tau e^{i\varphi} | -\alpha\rangle$
was studied in \cite{Brif1}.
The square of the distance between the coherent and cat states
with the same values of the parameter $\alpha$ equals
\begin{equation}
d^2\left(|\alpha;\varphi\rangle\,, |\alpha\rangle\right)=
\frac{1- \exp(-4|\alpha|^2)}{1+\cos\varphi \exp(-2|\alpha|^2)}.
 \label{dist-coh-cat}
 \end{equation}
For the distance from the vacuum state, we obtain
\begin{equation}
d^2\left(|\alpha;\varphi\rangle\,, | 0\rangle\right)=
\frac{2\left[1- \exp(-|\alpha|^2)\right]}{1+\cos\varphi \exp(-2|\alpha|^2)},
 \label{dist-vac-cat}
 \end{equation}
whereas the distance between two states with the same parameter $\alpha$
but different values of phases $\varphi_1$ and $\varphi_2$ reads
\begin{equation}
d^2\left(\varphi_1, \varphi_2\right)=
\frac{\left[1- \exp(-4|\alpha|^2)\right]
\left[1- \cos(\varphi_1- \varphi_2)\right]}
{\left[1+\cos\varphi_1 \exp(-2|\alpha|^2)\right]
\left[1+\cos\varphi_2 \exp(-2|\alpha|^2)\right]}.
\label{dist-cat-cat}
 \end{equation}
For $|\alpha|\gg 1$, we have
$d^2\left(\varphi_1, \varphi_2\right)\approx
2 \sin^2\left(\left|\varphi_1- \varphi_2\right|/2\right)$.

The $N$-distances between the same states have an extra factor $|\alpha|$:
\begin{equation}
d_N^2\left(|\alpha;\varphi\rangle\,, | 0\rangle\right)=|\alpha|^2
\frac{1- \cos\varphi\exp(-2|\alpha|^2)}{1+\cos\varphi \exp(-2|\alpha|^2)},
 \label{distN-vac-cat}
 \end{equation}
\begin{equation}
d_N^2\left(\varphi_1\,, \varphi_2\right)=
 \frac{|\alpha|^2 \left[1+ \exp(-4|\alpha|^2)\right]
\left[1- \cos(\varphi_1- \varphi_2)\right]}
{\left[1+\cos\varphi_1 \exp(-2|\alpha|^2)\right]
\left[1+\cos\varphi_2 \exp(-2|\alpha|^2)\right]}.
\label{distN-cat-cat}
 \end{equation}
Now we have
$d_N\approx \sqrt2 |\alpha| \sin\left(\left|\varphi_1-
\varphi_2\right|/2\right)$ for $|\alpha|\gg 1$.

Equations (\ref{dist-coh-cat})-(\ref{distN-cat-cat}) clearly show that the
YS-states are intermediate between even and odd ones. Moreover, we see that
the distance between the YS and the odd states with the same $|\alpha|$ is
greater than that between the YS and the even states, and the YS-state is
farther from the coherent state than the even state (whereas the Mandel
parameter does not distinguish the coherent and YS states at all).
This example demonstrates how the concept of distance helps to understand
better the properties of different families of quantum states and the mutual
relations between them.

\subsection{Coherent phase states}

As a further example, we consider the {\it coherent phase states\/}
\cite{Ler}
\begin{equation}
|\varepsilon\rangle =\sqrt{1-\varepsilon\varepsilon^*}\sum_{n=0}^\infty 
\varepsilon^n|n\rangle,
\qquad \hat{E}_-|\varepsilon\rangle = \varepsilon
|\varepsilon\rangle,
\qquad |\varepsilon|<1,
\label{eps-state}
\end{equation}
where
\[
\hat{E}_-\equiv\sum_{n=1}^\infty|n-1\rangle
\langle n| =\left(\hat{a}\hat{a}^{\dagger}\right)^{-1/2}\hat{a}
\]
is the Susskind-Glogower phase operator \cite{Suss} which can be considered
to certain extent as a quantum analogue of the classical phase
$e^{i\varphi}$.
The HS distance between the states $|\varepsilon_1\rangle$
and $|\varepsilon_2\rangle$ is given by
\begin{equation}
d(|\varepsilon_1\rangle ,|\varepsilon_2\rangle )
=\frac{\sqrt{2}|\varepsilon_1-\varepsilon_2|}
{|1-\varepsilon_1\varepsilon_2^*|}.
\label{eps-eps}
\end{equation}
It is proportional to the
geometric distance of the complex parameters $\varepsilon_1$ and 
$\varepsilon_2$ for $|\varepsilon_{1,2}|\ll 1$.
For any $|\varepsilon|<1$, the distance from the vacuum state is simply
$d(|\varepsilon\rangle ,|0\rangle )=\sqrt{2}\,|\varepsilon|$.
At the same time, the $d_N\,$-distance is given by
\begin{eqnarray}
d_{ N}^2\left(|\varepsilon_1\rangle,|\varepsilon_2\rangle \right) &=&
\frac{|\varepsilon_1|^2}{1-|\varepsilon_1|^2} +
\frac{|\varepsilon_2|^2}{1-|\varepsilon_2|^2}
\nonumber\\
&+&2\frac{
\left(1-|\varepsilon_1|^2\right)\left(1-|\varepsilon_2|^2\right)
\left[|\varepsilon_1\varepsilon_2|^2-
\mbox{Re}\left(\varepsilon_1\varepsilon_2^*\right)\right]}
{\left[1-2\mbox{Re}\left(\varepsilon_1\varepsilon_2^*\right)
+|\varepsilon_1\varepsilon_2|^2\right]^2}.
\label{dist-eps}
\end{eqnarray}
In particular,
$d_N(|\varepsilon\rangle ,|0\rangle )=
|\varepsilon|\left(1-|\varepsilon|^2\right)^{-1/2}$.

\subsection{Thermal states}

The pure quantum state (\ref{eps-state}) has the same probability
distribution $|\,\langle n|\varepsilon\rangle\,|^2$ as the
mixed {\it thermal state\/} described by the statistical operator
\begin{equation}
\hat{\rho} =\frac{1}{1+\overline{n}}\sum_{n=0}^\infty
\left(\frac{\overline{n}}{1+\overline{n}}\right)^n|n\rangle
\langle n|
\label{def-therm}
\end{equation}
provided that one identifies the mean photon number $\overline{n}$
with $|\varepsilon|^2/\left(1-|\varepsilon|^2\right)$ \cite{AhaLer}.
Moreover, the state (\ref{eps-state}) arises naturally as an exact solution
to some nonlinear modifications of the Schr\"odinger equation \cite{DoMi},
so it can be named also a ``pseudothermal state'' \cite{DoMi}.
Therefore it is interesting to compare the expressions (\ref{eps-eps})
and (\ref{dist-eps}) for the distances between ``pseudothermal'' states
with the analogous formulae for the true thermal states.

The HS distance between two states (\ref{def-therm}) reads
\begin{equation}
d^{(HS)}(\overline{n}_1,\overline{n}_2)
=\frac{\sqrt{2}\left|\overline{n}_1-\overline{n}_2 \right|}
{\sqrt{\left(1+2\overline{n}_1\right)\left(1+2\overline{n}_2
\right)\left(1+\overline{n}_1+\overline{n}_2\right)}}\,.
\label{therm-2}
\end{equation}
Although it is proportional to the difference of the mean photon numbers,
it goes to zero when $\overline{n}_{1,2}\to\infty$ and
$|\overline{n}_1 -\overline{n}_2|=\mbox{const}$.
The distance to the ground state equals
\begin{equation}
d^{(HS)}(\overline{n},0)=\frac{\overline{n}\,\sqrt{2}}
{\sqrt{(1+\overline{n})(1+2\overline{n})}}\,,
\end{equation}
and it tends to $1$ when $\overline{n}\to\infty$, i.e., to the value which is
$\sqrt2$ times less than the maximal possible Hilbert-Schmidt distance.
These results become clear if one remembers that
highly mixed states are located, in a sense, deeply ``inside'' the Hilbert
space, since the density operators form a convex set with the pure states
contained in the boundary \cite{Miel}.
Nonetheless, being justified from the mathematical point of view,
these properties do not agree completely with our
physical intuition, because usually we think on highly mixed
states as almost classical ones (all the coherence is lost), which must
be far away from the intrinsically quantum vacuum state. In particular,
it seems a little bit strange that high temperature states are closer
to the ground state than any pure Fock state.

Using the modified HS distance (\ref{d-root1})
(which coincides with the Bures-Uhlmann distance in the case involved)
we obtain
\begin{equation}
d^{(BU)}(\overline{n}_1,\overline{n}_2)=\sqrt{2}\left[1-
\frac{\sqrt{(1+\overline{n}_1)(1+\overline{n}_2)}+
\sqrt{\overline{n}_1\overline{n}_2}}
{1+\overline{n}_1+\overline{n}_2}\right]^{1/2}.
\label{BU-therm}
\end{equation}
In particular, the distance to the ground state equals
\begin{equation}
d^{(BU)}(\overline{n},0)=
\frac{\sqrt{2\overline{n}}}
{\left[\sqrt{1+\overline{n}}
\left(1+\sqrt{1+\overline{n}}\right)\right]^{1/2}}
\end{equation}
and it tends to the maximal possible value $\sqrt{2}$
when $\overline{n}\to\infty$.
It is interesting to compare this formula with the analogous one for the
``pseudothermal'' state (\ref{eps-state}), but written in terms of the mean
photon number:
\[
d^{(HS)}(|\varepsilon\rangle,0)=
\sqrt{\frac{2\overline{n}}{1+\overline{n}}}\;.
\]
We see that the BU-distance for the mixed states is always a little bit
less than the distance between the vacuum and the pure pseudothermal state
with the same value of $\overline{n}$, in agreement with the reasonings of
the preceding paragraph.
For $\overline{n}_{1,2}\gg 1$,
 (\ref{BU-therm}) is simplified
\begin{equation}
d^{(BU)}(\overline{n}_1,\overline{n}_2) \approx
\frac{\sqrt{2}\left|\sqrt{\overline{n}_1} -\sqrt{\overline{n}_1}\right|}
{\sqrt{\overline{n}_1+\overline{n}_2}}\,.
\label{BU-therm1}
\end{equation}

The square of the $N$-distance between two thermal states (\ref{def-therm})
reads
\begin{equation}
d_N(\overline{n}_1,\overline{n}_2) =
\frac{|\overline{n}_1-\overline{n}_2|\sqrt{
(1+\overline{n}_1+\overline{n}_2)^2
+2\overline{n}_1\overline{n}_2 (1+2\overline{n}_1)
(1+2\overline{n}_2)}}
{(1+2\overline{n}_1)(1+2\overline{n}_2)(1+\overline{n}_1+
\overline{n}_2)}.
\label{dist-plank-bad}
\end{equation}
As well as for the HS distance, the high temperature states
occur not very far from the ground state:
\[
d_N(\overline{n},0)=\frac{\overline{n}}{1+2\overline{n}} \;
\rightarrow\;\frac{1}{2}
\quad {\rm when} \quad \overline{n} \rightarrow \infty.
\]
At the same time, using the modified $N$-distance (\ref{tild-Zlob})
we obtain the expression
\begin{equation}
\tilde{d}_N^2(\overline{n}_1,\overline{n}_2)=
\overline{n}_1+\overline{n}_2-2\sqrt{\overline{n}_1\overline{n}_2}
\left(\frac{\sqrt{(1+\overline{n}_1)(1+\overline{n}_2)}+
\sqrt{\overline{n}_1\overline{n}_2}}
{1+\overline{n}_1+\overline{n}_2}\right)^2,
\label{dist-plank}
\end{equation}
which yields $\tilde{d}_N(\overline{n},0)=\overline{n}^{1/2}$,
as well as for pure states.
Analyzing formula (\ref{dist-eps}) for the $N$-distance between the
``pseudothermal'' states, one can check that the right-hand side attains the
minimum (for fixed absolute values $|\varepsilon_{1,2}|$) if
$\mbox{Re}\left(\varepsilon_1^*\varepsilon_2\right)=
|\varepsilon_1\varepsilon_2|$. This minimal distance can be written in
terms of $\overline{n}_{1,2}$ in the form very similar to (\ref{dist-plank}),
but the last factor has the exponent $3$ instead of $2$:
\begin{equation}
\tilde{d}_{N\min}^2(|\varepsilon_1\rangle, |\varepsilon_2\rangle)=
\overline{n}_1+\overline{n}_2-2\sqrt{\overline{n}_1\overline{n}_2}
\left(\frac{\sqrt{(1+\overline{n}_1)(1+\overline{n}_2)}+
\sqrt{\overline{n}_1\overline{n}_2}}
{1+\overline{n}_1+\overline{n}_2}\right)^3.
\label{dist-mineps}
\end{equation}
Since the fraction inside the parentheses does not exceed $1$ (this is
a consequence of the inequality
$\overline{n}_1+\overline{n}_2 \ge 2\sqrt{\overline{n}_1\overline{n}_2}\;$),
we have
$\tilde{d}_{N\min}(|\varepsilon_1\rangle, |\varepsilon_2\rangle) \ge
\tilde{d}_N(\overline{n}_1,\overline{n}_2)$
for any pair of pure and mixed
states with the same mean photon numbers.
Equations (\ref{dist-plank}) and (\ref{dist-mineps}) can be simplified for
$\overline{n}_{1,2} \gg 1$:
\[
\tilde{d}_N^2(\overline{n}_1,\overline{n}_2) \approx
\overline{n}_1+\overline{n}_2 -
\frac{8\left( \overline{n}_1 \overline{n}_2 \right)^{3/2}}
{\left( \overline{n}_1+\overline{n}_2\right)^2}
\]
\[
\tilde{d}_{Nmin}^2(|\varepsilon_1\rangle, |\varepsilon_2\rangle) \approx
\overline{n}_1+\overline{n}_2 -
\frac{16\left( \overline{n}_1 \overline{n}_2 \right)^{2}}
{\left( \overline{n}_1+\overline{n}_2\right)^3}.
\]
If also
$\left|\overline{n}_1-\overline{n}_2\right| \ll
\overline{n}_{1,2}\;$,
then we obtain approximate expressions resembling formula (\ref{dist-n*})
for the {\it quasi\/}distance between the Fock states, but with different
coefficients
\[
\tilde{d}_N^2(\overline{n}_1,\overline{n}_2) \approx
\sqrt3\left|\sqrt{\overline{n}_1}-\sqrt{\overline{n}_2}\right|=
\frac{\sqrt3\left|\overline{n}_1-\overline{n}_2\right|}
{\sqrt{\overline{n}_1}+\sqrt{\overline{n}_2}},
\]
\[
\tilde{d}_{N\min}(|\varepsilon_1\rangle, |\varepsilon_2\rangle) \approx
2\left|\sqrt{\overline{n}_1}-\sqrt{\overline{n}_2}\right|=
\frac{2\left|\overline{n}_1-\overline{n}_2\right|}
{\sqrt{\overline{n}_1}+\sqrt{\overline{n}_2}}.
\]
The dependences of different distances between the vacuum and thermal
or ``pseudothermal'' states on the mean photon number $\overline{n}$ are
shown in Fig. 2.
The distances of the pure states are larger than analogous distances
of the mixed states with the same mean photon numbers, excepting the
case of the $\tilde{d}_N\;$-distance, which is the same both for the thermal
and the phase coherent states.
We can conclude that the $\tilde{d}_N\,$-distance seems to be the most
adequate from the physical point of view (at least for thermal states).

\section{``Classical-like'' quantum distances}\label{dist-clas}
\setcounter{equation}{0}

It is accepted that quantum states are described in terms of the
 wave functions (state vectors in the Hilbert space)
or density matrices (statistical operators).
However, these {\it complex-valued\/} objects have rather indirect
relations to the results of measurements, which are expressed in terms
of real positive probabilities.
Recently, a new formulation of quantum mechanics in terms of positive
classical probability distributions was proposed
\cite{mancini3-mancini4,voljrlr,bregenz}. It is
a natural consequence of the concepts of the so-called {\it
symplectic tomography\/} developed in \cite{mancini1,mancini2}.

Let us introduce the two-parameter
family of quadrature operators $\hat X_{\mu\nu}=\mu\hat q +\nu\hat p$,
$-\infty<\mu,\nu<\infty$, where $\hat q$ and $\hat p$
are the usual coordinate and momentum operators (in one dimension,
for simplicity).
It can be shown
that
the probability distribution $w_{\mu\nu}\left(X\right)$ of the real
eigenvalues of the Hermitian operator $\hat X_{\mu\nu}$
is given by the following integral transform of the Wigner function:
\begin{equation}\label{w}
w_{\mu\nu}(X)=\int \frac{dq\, dp}{2\pi}\,
\delta(\mu q + \nu p -X) W(q,p).
\end{equation}
The reciprocal transform
\begin{equation}
W(q,p)= \frac{1}{2\pi}\int dX\, d\mu\, d\nu\,
\exp[i(X-\mu q -\nu p)] w_{\mu\nu}(X)
\label{recipr}
\end{equation}
enables to express any Wigner function (and, consequently, any density
matrix) in terms of the positive marginal
probability distributions $w_{\mu\nu}(X)$ which can be obtained, in
principle, directly from an experiment with the aid of the homodyne
detection schemes. Consequently, the description in terms of the family
of classical distributions $w_{\mu\nu}(X)$ is completely equivalent to the
standard description in terms of the density matrix or the wave function.
This fact is the basis of the ``classical-like'' formulation of quantum
mechanics \cite{mancini3-mancini4,voljrlr,bregenz,olgajetf,mancini5,spin}.
In this formulation, every quantum state is described not by a single
complex-valued function $\psi(x)$ or $\rho(x,x')$, but by an infinite set
of {\it classical positive probability distributions\/} $w_{\mu\nu}(X )$,
$-\infty<\mu,\nu<\infty$.
For example, the Fock state of the harmonic oscillator $| n\rangle$
is described by the family of the marginal distributions~\cite{voljrlr}
\begin{equation}\label{M3}
w_{\mu\nu}^{(n)}\left(X\right)=
w_{\mu\nu}^{(0)}\left(X\right)\frac {1}{2^nn!}
H_n^2\left(\frac {X}{\sqrt {\mu^2+\nu^2}}\right),
\end{equation}
where $H_n(z)$ is the Hermite polynomial, while the marginal distribution
$w_{\mu\nu}^{(0)}(X)$ of the vacuum state reads
\begin{equation}\label{M4}
w_{\mu\nu}^{(0)}\left(X\right)=\frac {1}{\sqrt {\pi(\mu^2+\nu^2)}}
\exp\left(-\frac {X^2}{\mu^2+\nu^2}\right).
\end{equation}

Now, considering the quantum states described by two different sets of the
marginal distributions $w_{\mu\nu}^{(a)}(X)$ and $w_{\mu\nu}^{(b)}(X)$
we can define the ``classical-like'' distance between these states as
\begin{equation}
{\cal D}_{ab}^{{\cal C}}=\int d\mu\, d\nu\, g(\mu,\nu)
d_{ab}^{{\cal C}}\left(w_{\mu\nu}^{(a)}, w_{\mu\nu}^{(b)}\right),
\label{def-clasdis}
\end{equation}
where
$d_{ab}^{{\cal C}}\left(w_{\mu\nu}^{(a)}, w_{\mu\nu}^{(b)}\right)$ is
some {\it classical\/} distance between the distributions
$w_{\mu\nu}^{(a)}(X)$ and $w_{\mu\nu}^{(b)}(X)$. A positive weight function
$g(\mu,\nu)$ is introduced to ensure the convergence of the
integral over $\mu,\nu$. Evidently, if the ``partial distance''
$d_{ab}^{{\cal C}}\left(w_{\mu\nu}^{(a)}, w_{\mu\nu}^{(b)}\right)$
satisfies the triangle inequality for all fixed values of $\mu,\nu$, this
inequality remains valid after multiplying by the positive function
$g(\mu,\nu)$ and the subsequent integration over $\mu,\nu$.

Let us consider, for example, the ``Kakutani-Hellinger-Matusita distance''
\cite{handbook,jacod}
between two real nonnegative distributions
$P_1\left (x\right )$ and $P_2\left (x\right )$
\begin{equation}\label{dist-1}
d_{{\cal H}}\left(P_1, P_2\right)=
\left [ \int dx\left (\sqrt {P_1\left (x\right )}-
\sqrt {P_2\left (x\right )}\right )^2\right ]^{1/2}.
\end{equation}
Taking into account the normalization condition
we arrive at the ``classical-like'' analogue of the Bures-Uhlmann distance
\begin{equation}
{\cal D}_{ab}^{{\cal H}}=\sqrt2\int d\mu\, d\nu\, g(\mu,\nu)\left[1-
\int dX \sqrt{w_{\mu\nu}^{(a)}(X) w_{\mu\nu}^{(b)}(X)}\,
\right]^{1/2}.
\label{clasdis}
\end{equation}
The ``classical-like'' analogue of the JMG-distance (\ref{d-jauch-dieks})
is obtained if one chooses for
$d_{ab}^{{\cal C}}$
the classical Kolmogorov distance \cite{handbook}
\begin{equation}
d_{{\cal K}}\left(P_1, P_2\right)= \int dx\left| P_1(x)-P_2(x)\right|.
\label{Kolm}
\end{equation}

To illustrate the new approach, let us consider the
${\cal D}^{{\cal H}}\,$-distance (\ref{clasdis}) between two coherent
states $|\alpha\rangle$ and $|\beta\rangle$. Each of these states is
described by means of the families of the marginal distributions like
\begin{equation}\label{M2exam}
w_{\mu \nu }^{(\alpha)}(X)=\frac {1}{\sqrt {\pi \left(\mu^2 +\nu^2\right)}}
\exp\left(-\frac {\left[X-\overline X_{\alpha}(\mu,\nu)\right]^2}
{\mu^2 +\nu^2}\right),
\end{equation}
\[
\overline X_{\alpha}(\mu,\nu)=\sqrt2 \left(\mu \,\mbox {Re}\,\alpha
+ \nu \,\mbox {Im}\,\alpha\right) .
\]
Introducing the polar coordinates in the $\mu\,\nu$ plane,
$\mu=R\cos\vartheta$, $\nu=R\sin\vartheta$, we see that the
${\cal D}^{{\cal H}}\,$-distance between the
coherent states depends on $|\alpha-\beta|$ only:
\begin{equation}
{\cal D}_{\alpha\beta}^{{\cal H}}=\int_0^{\infty} R\, dR
\int_0^{2\pi}d\vartheta \,g(R,\vartheta)
\left\{
2-2\exp\left[-\frac12 |\alpha-\beta|^2\cos^2(\vartheta-\varphi)\right]
\right\}^{1/2}
\label{heldiscoh}
\end{equation}
(here $\varphi$ is the phase of the complex number $\alpha-\beta$).
It is convenient to choose the weight function $g(R,\vartheta)$ independent
on $\vartheta$ and to impose the condition $\int_0^{\infty}g(R)R\,dR=1$.
Then for close coherent states, we have
${\cal D}_{\alpha\beta}^{{\cal H}}=4|\alpha-\beta|$ if $|\alpha-\beta|\ll 1$.
When $|\alpha-\beta|\to\infty$, the ${\cal D}^{{\cal H}}\,$-distance
tends to the constant value $2\pi\sqrt2$.

The integral over $\mu,\nu$ can be calculated explicitly for classical-like
 {\it distinguishability measures\/} (DM)
which are defined by the
same formula (\ref{def-clasdis}) but without imposing the requirement
(\ref{treug}) (the triangle inequality) on the function
$d_{ab}^{{\cal C}}\left(w_{\mu\nu}^{(a)}, w_{\mu\nu}^{(b)}\right)$.
The distinguishability measures are frequently used
in the classical statistics and the information theory \cite{handbook}.
Their applications to quantum mechanical problems were discussed recently
in \cite{ved,fuchs}.
The most known examples of classical DM are the
{\it Bhattacharyya coefficient} \cite{handbook}
\begin{equation}
{\cal B}\left(P_1, P_2\right)=
-\ln\int dx\sqrt{P_1(x)P_2(x)}
\label{Bhat}
\end{equation}
and the {\it Kullback-Liebler distinguishability measure}
\cite{handbook}
\begin{equation}
{\cal J}\left(P_1, P_2\right)=
\int dx\left[ P_1(x)-P_2(x)\right]\ln\frac{P_1(x)}{P_2(x)}.
\label{Kull}
\end{equation}
For coherent states, both these measures yield similar dependences on the
parameters $\alpha$ and $\beta$, which differ only in a scale factor
(we assume the same weight function $g(\mu,\nu)$ as above):
\begin{equation}
{\cal D}_{\alpha\beta}^{({\cal J})}= 8{\cal D}_{\alpha\beta}^{({\cal B})}=
4\pi |\alpha-\beta|^2.
\label{BJdiscoh}
\end{equation}
These quantum DM are unbounded when $|\alpha-\beta|\to\infty$, but they
do not satisfy the triangle inequality.

\section{Conclusion}\label{concl}

Let us summarise the main results of the paper.
We have obtained new inequalities for the Hilbert-Schmidt distance and its
modifications, which
can be used for evaluating the ``degree of proximity''
between close quantum states.
We have given new expressions for the Hilbert-Schmidt distance in terms of
quasiprobability distributions and in terms of the ordered moments.
We have constructed the distances
which are sensitive to the energy of quantum states. These
``$N$-distances'' are unlimited and they distinguish different orthogonal
states.
Besides, we have shown how the concept of distance can be introduced
in the framework of the new ``classical-like'' formulation of quantum
mechanics in terms of positive probability distributions of the rotated
(in the phase space) quadrature operators.

\subsection*{Acknowledgements}
VVD, OVM, and VIM are grateful to the Arbeitsgruppe
``Nichtklassische Strahlung,'' der Max-Planck-Gesellschaft for the
hospitality during their visits to Berlin.
OVM and VIM acknowledge a partial support of the Russian Foundation for
Basic Research under the Project~~96-02-17222.

\newpage
\begin{figure}
\caption{The dependences  of  the $N$-distance (three upper curves) and
the Hilbert-Schmidt distance (three lower curves) between the coherent
state $|\alpha\rangle$ and the Fock states $|m\rangle$ with $m=1,2,3$,
versus the mean photon number in the coherent state $|\alpha|^2$.
The order of curves from bottom to top (in the part of plot nearby the
vertical axis): the lower curves correspond to $m=1$ while the upper
ones correspond to $m=3$.
}
\end{figure}

\begin{figure}
\caption{Different distances between the vacuum  and
the thermal (mixed) and pseudothermal (pure phase coherent) states versus
the mean photon number.
The order of the curves in the right-hand side of the plot
(from bottom to top):
$N$-distance for the thermal state;
the Hilbert-Schmidt distance for the thermal state;
the Bures-Uhlmann distance for the thermal state;
the Hilbert-Schmidt distance for the pseudothermal state;
$N$-distance for the pseudothermal state (it coincides with the modified
$N$-distance $\tilde{d}_N$ for the thermal state in the case concerned).
}
\end{figure}

\newpage
\begin{thebibliography}{99}

\bibitem{glauber63} Glauber, R. J.,  Phys. Rev. {\bf 131}, 2766 (1963).

\bibitem{DMM-74} Dodonov,~V.~V., Malkin,~I.~A., and Man`ko,~V.~I.,
 Physica {\bf 72},  597 (1974).

\bibitem{titgla} Titulaer,~U.~M. and Glauber,~R.~J.,  Phys. Rev.
{\bf 145}, 1041 (1966).

\bibitem{zofia} Bialynicki-Birula,~Z., Phys. Rev. {\bf 173}, 1207 (1968).

\bibitem{YuS} Yurke,~B. and Stoler,~D., Phys. Rev. Lett. {\bf 57}, 13 (1986).

\bibitem{mancini3-mancini4}
Mancini,~S., Man'ko,~V.~I., and Tombesi,~P., Phys. Lett. A  {\bf 213},  1
(1996); Found. Phys. {\bf 27}, 801 (1997).

\bibitem{Barg} Bargmann,~V.,  Ann. Math. {\bf 59}, 1 (1954).
\bibitem{Baltz} von Baltz,~R., Europ. J. Phys. {\bf 11}, 215 (1990).
\bibitem{Anand91} Anandan,~J., Found. Phys. {\bf 21}, 1265 (1991).
\bibitem{Pati91} Pati,~A.~K., Phys. Lett. A {\bf 159}, 105 (1991).
\bibitem{AnAh90} Anandan,~J. and Aharonov,~Y., Phys. Rev. Lett.
{\bf 65}, 1697 (1990).
\bibitem{Mont} Montgomery,~R., Comm. Math. Phys. {\bf 128}, 565 (1990).
\bibitem{Pati92} Pati,~A.~K., J. Phys. A {\bf 25}, L1001 (1992).
\bibitem{Grig} Grigorenko,~A.~N., Phys. Rev. A {\bf 46}, 7292 (1992).
\bibitem{Hub93-1} H\"ubner,~M., Phys. Lett. A {\bf 179},  221 (1993).
\bibitem{HirHam} Hirayama,~M. and Hamada,~T., Prog. Theor. Phys.
{\bf 91}, 991 (1994).
\bibitem{BraMil} Braunstein,~S.~L. and Milburn,~G.~J.,
 Phys. Rev. A {\bf 51}, 1820 (1995).
\bibitem{Provost} Provost,~J.~P. and Vallee,~G., Comm. Math. Phys.
{\bf 76}, 289 (1980).
\bibitem{Page} Page,~D.~N., Phys. Rev. A {\bf 36}, 3479 (1987).
\bibitem{Anan90} Anandan,~J., Phys. Lett. A {\bf 147}, 3 (1990).
\bibitem{Trif} Trifonov,~D.~A., J. Math. Phys. {\bf 34}, 100 (1993).
\bibitem{Abe} Abe,~S., Phys. Rev. A {\bf 48}, 4102 (1993).

\bibitem{Woot} Wootters,~W.~K., Phys. Rev. D {\bf 23}, 357 (1981).
\bibitem{Braun} Braunstein,~S.~L. and Caves,~C.~M., Phys. Rev. Lett.
{\bf 72},  3439 (1994).
\bibitem{Plast} Ravicul\'e,~M., Casas,~M., and Plastino,~A., Phys. Rev.
{\bf 55}, 1695 (1997).
\bibitem{Jauch} Jauch,~J.~M., Misra,~B., and Gibson,~A.~G., Helv. Phys. Acta
{\bf 41}, 513 (1968).
\bibitem{Dieks} Dieks,~D. and Veltkamp,~P., Phys. Lett. A {\bf 97}, 24 (1983).
\bibitem{Hil} Hillery,~M., Phys. Rev. A {\bf 35}, 725 (1987);
{\bf 39}, 2994 (1989).
\bibitem{Ruch}
Ruch,~E., Theor. Chim. Acta {\bf 38}, 167 (1975);
Schranner,~R., Seligman,~T.~H., and Ruch,~E., J. Chem. Phys. {\bf 69}, 386
(1978);
Lesche,~B. and Ruch,~E., J. Chem. Phys. {\bf 69}, 393 (1978);
Busch,~P. and Ruch,~E., Int. J. Quant. Chem. {\bf 41},  163 (1992).
\bibitem{Guz} Caianiello,~E.~R. and Guz,~W., Phys. Lett. A {\bf 126}, 223
(1988).
\bibitem{Bures} Bures,~D., Trans. Am. Math. Soc. {\bf 135}, 199 (1969).
\bibitem{Uhlm} Uhlmann,~A., Rep. Math. Phys. {\bf 9},  273 (1976).
\bibitem{Gud} Gudder,~S., Marchand,~J.-P., and Wyss,~W.,  J. Math. Phys.
{\bf 20}, 1963 (1979).
\bibitem{Hub} H\"ubner,~M., Phys. Lett. A {\bf 163},  239 (1992).
\bibitem{Joz} Jozsa,~R., J. Mod. Opt. {\bf 41}, 2315 (1994).
\bibitem{Hub93-2} H\"ubner,~M., Phys. Lett. A {\bf 179},  226 (1993).
\bibitem{Slat1} Slater, P. B., J. Phys. A {\bf 29}, L271 (1996).
\bibitem{Twam} Twamley,~J., J. Phys. A {\bf 29}, 3723 (1996).
\bibitem{Slat2} Slater, P. B., J. Phys. A {\bf 29}, L601 (1996).
\bibitem{scut} Paraoanu, Gh.-S. and Scutaru, H., Los Alamos
 Report quant-ph/9703051.
\bibitem{Sam} Samuel,~J. and Bhandari,~R., Phys. Rev. Lett. {\bf 60}, 2339
(1988).
\bibitem{Wun} W\"unsche,~A., Appl. Phys. B {\bf 60},  S119 (1995).
\bibitem{Orl} Kn\"oll,~L. and Orlowski,~A., Phys. Rev. A {\bf 51}, 1622
(1995).

\bibitem{cahgla}
Cahill,~K.~E. and Glauber,~R.~J., Phys. Rev.  {\bf 177}   1882 (1969).
\bibitem{wigner32}
Wigner,~E., Phys. Rev.  {\bf 40},   749 (1932).

\bibitem{husimi-kano}
Husimi,~K., Proc. Phys. Math. Soc. Jpn  {\bf 23},  264 (1940);
Kano,~Y., J. Math. Phys.  {\bf 6},   1913 (1965).

\bibitem{glauber-sudarshan}
Glauber,~R.~J., Phys. Rev. Lett.  {\bf 10},   84 (1963);
Sudarshan,~E.~C.~G., Phys. Rev. Lett.  {\bf 10},   277 (1963).
\bibitem{wu} W\"unsche,~A., Quantum Opt. {\bf 2}, 453 (1990).

\bibitem{wubu} W\"unsche,~A. and Bu\v{z}ek,~V., Quantum Semiclass. Opt.
{\bf 9}, 631 (1997).
\bibitem{wuensche}
 W\"unsche,~A., J. Mod. Opt. {\bf 44}, 2293 (1997).

\bibitem{QE} Dodonov,~V.~V., Man'ko,~V.~I., and Rudenko,~V.~N., Kvantov.
\'Elektron. {\bf 7}, 2124 (1980)  [Sov. J. Quantum Electron. {\bf 10}, 1232
(1980)].
\bibitem{Sch} Schleich,~W., Walls,~D.~F., and Wheeler,~J.~A.,
 Phys. Rev. A {\bf 38},  1177 (1988)
\bibitem{squeez} Stoler,~D., Phys. Rev. D {\bf 1}, 3217 (1970);
Yuen,~H.~P., Phys. Rev. A {\bf 13}, 2226 (1976).

\bibitem{ManW} Man'ko,~V.~I. and W\"unsche,~A., Quantum Semiclass. Opt.
{\bf 9}, 381 (1997).
\bibitem{Brif1} Brif,~C., Ann. Phys. (NY) {\bf 251}, 180 (1996).
\bibitem{Ler} Lerner,~E.~C., Huang,~H.~W., and Walters,~G.~E.,
 J. Math. Phys. {\bf 11}, 1679 (1970);
Ifantis,~E.~K., J. Math. Phys. {\bf 13}, 568 (1972);
Shapiro,~J.~H. and Shepard,~S.~R., Phys. Rev. A {\bf 43}, 3795 (1991);
 Brif,~C., Quantum Semiclass. Opt. {\bf 7}, 803 (1995).
\bibitem{Suss} Susskind, L. and Glogower, J., Physics {\bf 1}, 49 (1964);
Carruthers, P. and Nieto, M., Rev. Mod. Phys. {\bf 40}, 411 (1968);
Loudon, R., ``The Quantum Theory of Light'' (Clarendon, Oxford 1973).
\bibitem{AhaLer} Aharonov,~Y., Lerner,~E.~C., Huang,~H.~W., and Knight,~J.~M.,
J. Math. Phys. {\bf 14}, 746 (1973).
\bibitem{DoMi} Dodonov,~V.~V. and Mizrahi,~S.~S., Ann. Phys. (NY)
{\bf 237}, 226 (1995).
\bibitem{Miel} Mielnik,~B., Comm. Math. Phys. {\bf 37}, 221 (1974).

\bibitem{voljrlr}
Man'ko,~V.~I., J. Russ. Laser Res. (Plenum)  {\bf 17}, 579 (1996).
%
\bibitem{bregenz}
Man'ko,~V.~I., ``Symmetries in Science IX'' (Edited by B.~Gruber and
M.~Ramek) (Plenum, New York 1997), p.~215.

\bibitem{mancini1}
Mancini,~S., Man'ko,~V.~I., and Tombesi,~P., Quantum Semiclass. Opt.
 {\bf 7}, 615 (1995).

\bibitem{mancini2}
D'Ariano,~G.~M., Mancini,~S., Man'ko,~V.~I., and Tombesi,~P.,
 Quantum Semiclass. Opt.  {\bf 8},   1017 (1996).


\bibitem{olgajetf}
Man'ko,~V.~I. and Man'ko,~O.~V., Zhurn. \'Eksp. Teor. Fiz. {\bf 112}, 796
(1997) [JETP {\bf 85}, 430 (1997)];
 J. Russ. Laser Res. (Plenum) {\bf 18}, 411 (1997);
Man'ko,~V.~I. and Safonov,~S.~S., J. Russ. Laser Res. (Plenum)
{\bf 18}, 537 (1997); Theor. Math. Phys. {\bf 112}, 1172 (1997).

\bibitem{mancini5}
Mancini,~S., Man'ko,~V.~I., and Tombesi,~P., Europhys. Lett. {\bf 37}, 79
(1997); J. Mod. Opt. {\bf 44}, 2281 (1997).

\bibitem{spin}
Dodonov,~V.~V. and Man'ko,~V.~I., Phys. Lett. A {\bf 229}, 335 (1997).

\bibitem{handbook} Ben-Bassat,~M., ``Classification,
Pattern Recognition, and Reduction of Dimensionality''
(Handbook of Statistics, vol. 2).
(Edited by  P.~R.~Krishnaiah and L.~N.~Kanal)
(North-Holland, Amsterdam 1982), p. 773.
\bibitem{jacod}
Shiryayev,~A.~N., ``Probability'' (Springer, Berlin 1984);
Jacod,~J. and Shiryaev,~A.~N., ``Limit Theorems for Stochastic Processes''
(Grundlehren der mathematischen Wissenschaften, vol.~288,
A Series of Comprehensive Studies in Mathematics) (Springer, Berlin 1987).

\bibitem{ved} Vedral,~V., Plenio,~M.~B., Rippin,~M.~A., and Knight,~P.~L.,
 Phys. Rev. Lett. {\bf 78}, 2275 (1997).
\bibitem{fuchs} Fuchs, C. A. and van de Graaf, J.,
 Los Alamos Report quant-ph/9712042.

\end{thebibliography}
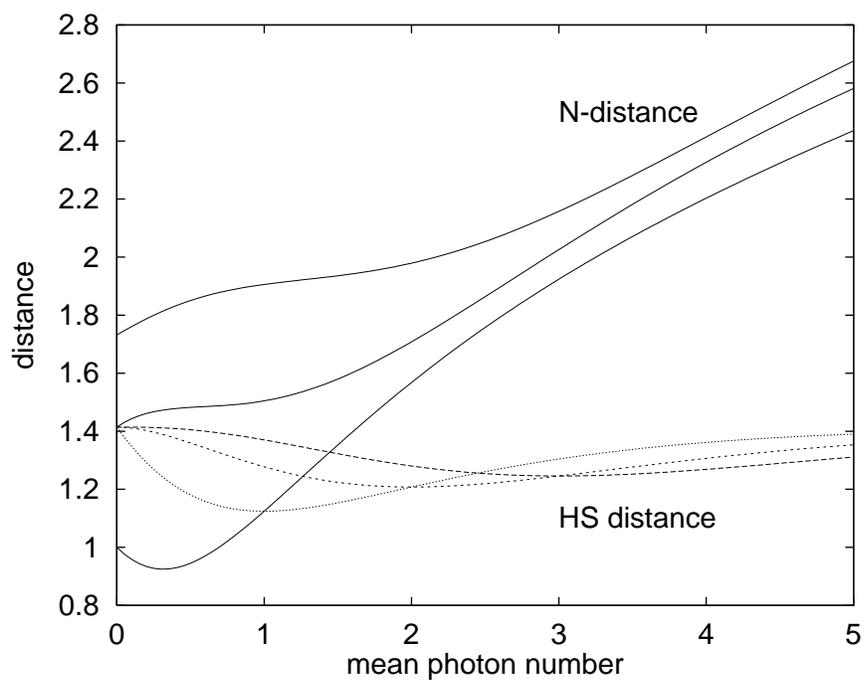
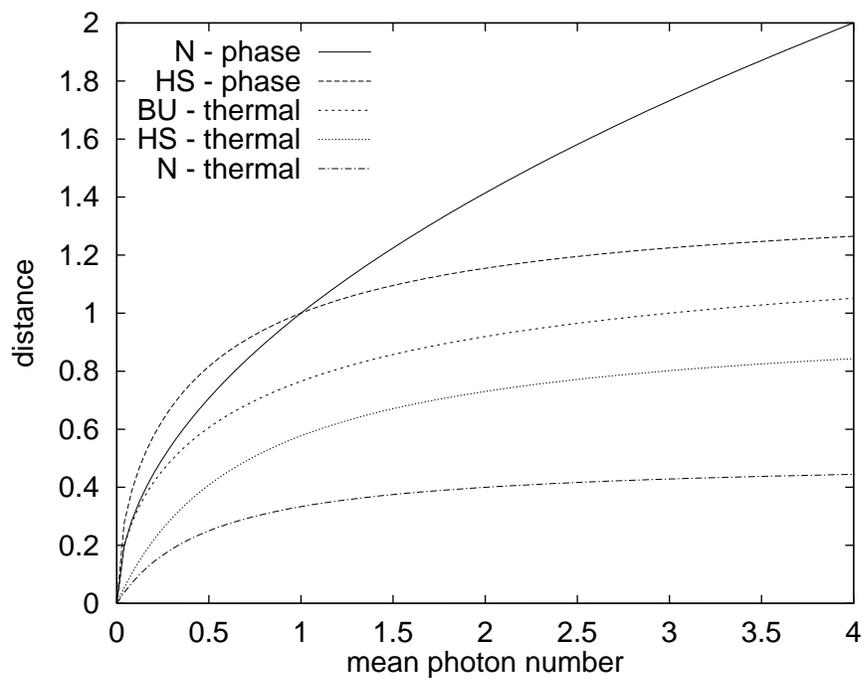
\end{document}